\documentclass[twocolumn,nobibnotes,aps,pra,reprint,superscriptaddress]{revtex4-1}
\usepackage{inputenc}
\usepackage[T1]{fontenc}
\usepackage{lmodern}
\usepackage{amsmath,amsfonts,amssymb,amstext,amscd,amsthm,dsfont}
\usepackage[mathscr]{eucal}
\usepackage{mathtools,bbm}
\usepackage{afterpage}
\usepackage{calrsfs,enumerate}
\usepackage[colorlinks=true,citecolor=blue,linkcolor=blue,urlcolor=blue]{hyperref}
\usepackage[normalem]{ulem}
\usepackage{graphicx}
\usepackage{latexsym}
\usepackage{bm}
\usepackage{makeidx}
\usepackage{subfigure}
\usepackage{bbold}
\usepackage{tikz}
\usetikzlibrary{arrows,matrix,positioning}
\usepackage{color}
\newcommand{\mathd}{\mathrm{d}}
\newcommand\ket[1]{\left|#1\right>}
\newcommand\bra[1]{\left<#1\right|}

\newcommand\projj[2]{\left|#1\right>\!\left<#2\right|}
\newcommand\proj[1]{\left|#1\right>\!\left<#1\right|}

\newcommand{\psp}{\phantom{\varphi_{\psi(t),i}}\hspace{-0.95cm}}

\newcommand{\reff}[1]{Eq.~\eqref{#1}}  
\newcommand{\reffs}[1]{Eqs.~\eqref{#1}}  

\def\bf{\textbf}

\begin{document}
%
%
%
%
%
\title{Rate Operator Unravelling for Open Quantum System Dynamics}

\author{Andrea Smirne}
\affiliation{Dipartimento di Fisica ``Aldo Pontremoli'', Universit{\`a} degli Studi di Milano, e Istituto Nazionale di Fisica Nucleare, Sezione di Milano, Via Celoria 16, I-20133 Milan, Italy}
\affiliation{Institute of Theoretical Physics, Universit{\"a}t Ulm, Albert-Einstein-Allee 11D-89069 Ulm, Germany}

\author{Matteo Caiaffa}
\affiliation{SUPA and Department of Physics, University of Strathclyde, Glasgow G4 0NG, UK}

\author{Jyrki Piilo}
\affiliation{QTF Centre of Excellence, Turku Centre for Quantum Physics, Department of Physics and Astronomy, University of Turku, FI-20014, Turun Yliopisto, Finland}

\begin{abstract}
Stochastic methods with quantum jumps are often used to solve open quantum system dynamics. 
Moreover, they provide insight into fundamental topics,
as the role of measurements in quantum mechanics 
and the description of non-Markovian memory effects. However, there is no unified framework to use quantum jumps to describe open system dynamics
in any regime. We solve this issue by developing the Rate Operator Quantum Jump (ROQJ) approach. The method not only applies to both Markovian and non-Markovian evolutions, but also allows us
to unravel master equations for which previous methods do not work.  
In addition, ROQJ yields a rigorous measurement-scheme interpretation for a
wide class of dynamics, including a
set of master equations with negative decay rates, and sheds light on different types of memory effects which arise when using stochastic quantum jump methods.
\end{abstract}
\maketitle
\textit{Introduction.---}%
Any realistic description of a quantum system should take into account
its interaction with the surrounding environment \cite{Breuer2002,Rivas2012}. 
Many different approaches 
have thus been developed to characterize the evolution of open quantum systems,
ideally covering different models and regimes, yet keeping the degree of complexity 
manageable \cite{Tanimura1989,Macri1992,Garraway1997,Breuer2008,Prior2010,Zhang2012,Gualdi2013,Ciccarello2013,Diosi2014,Vacchini2016,Tamascelli2018,Strathearn2018,Lambert2019,Luchnikov2019}.

Quantum unravellings yield a practically and conceptually useful tool,
mapping a given master equation to one (of the infinitely many possible)
pure-state stochastic evolution, which reproduces the given master equation on average \cite{Carmichael1993,Gardiner2004}.
On the one hand,
this leads to a linear scaling of the simulation cost with the Hilbert
space dimension of the open system, instead of the quadratic scaling which
would affect the direct integration of the master equation. 
On the other hand, unravellings might provide us with a clear physical picture of
the environmental influence on the open system evolution.
In particular, the stochastic pure-state evolution can be seen as the result of
a continuous measurement operated on the open system,
so that the master equation would correspond to the continuous action 
of a non-selective observer (the environment). 
This is the case, for example,
in the well-known Monte Carlo wave function (MCWF) method \cite{Dalibard1992,Plenio1998},
where the open-system pure state is subjected to a deterministic
evolution interrupted by random and discontinuous jumps.
Such piecewise deterministic evolutions under continuous monitoring
have been observed in several experimental platforms \cite{Basche1995,Peil1999,Jelezko2002,Gleyzes2007,Vijay2011,Minev2019}.

Memory effects pose some relevant challenges to 
unravelling methods, so that novel strategies need to be developed
to deal with non-Markovian dynamics \cite{Diosi1997,Breuer2004,Piilo2008,Barchielli2010,Suess2014,Gasbarri2018,Megier2018}.
Many non-equivalent definitions
have been introduced \cite{Breuer2009,Rivas2010,Li2018}, but
broadly speaking we can say that non-Markovian dynamics are characterized by a two-fold 
exchange of information between the open system and the environment, which
leads to memory effects and,
from the mathematical point of view, breaks relevant divisibility properties
of the dynamical maps fixing the open-system evolution.
The non-Markovian quantum jump (NMQJ) approach \cite{Piilo2008,Piilo2009}
accounts for the information flowing back to the open system 
by means of reversed jumps, which generalize the quantum jumps of the MCWF.
However, it is not clear to what extent, if at all, the continuous-measurement interpretation can be extended
to this and the other non-Markovian unravellings \cite{Gambetta2003,Diosi2008}. The basic intuition is that the (continuous)
measurements would affect in a non-trivial way the back-flow of information
to the open system and hence the subsequent dynamics, thus generating an evolution
which is not the same as the one given by the master equation to be unravelled \cite{Gambetta2003,Li2018}.

Here, first we show that a fully consistent continuous-measurement interpretation \cite{Barchielli1991}
can be formulated for any positive (P)-divisible dynamics \cite{Vacchini2011,Chruscinski2014,Wissmann2015},
via a jump unravelling approach which relies on the
diagonalization of a proper rate operator, and is named rate operator quantum jump (ROQJ).
The class of P-divisible dynamics includes master equations with negative rates
and is
larger than the one where MCWF applies,
thus highlighting the subtle border between Markovianity and non-Markovianity within the context of quantum
unravellings.
Furthermore, we extend ROQJ to deal with any open-system dynamics, including those
where at least one master-equation coefficient is negative from the very beginning of the evolution \cite{Cresser2010,Hall2014,Bernardes2015,Megier2017,Ferialdi2017},
so that other non-Markovian techniques, such as NMQJ, cannot be used.

\paragraph*{Quantum jumps for P-divisible dynamics.---}%
As usual within the unravelling methods, we start from
the master equation describing the dynamics
of the open quantum system of interest.
Any trace and Hermiticity preserving (time-local) master equation 
$\mathd \rho(t)/\mathd t = \mathcal{L}_t[{\rho(t)}]$ for the open-system state $\rho(t)$
can be written as~\cite{Gorini1976}
\begin{eqnarray}\label{eq:me}
&&\mathcal{L}_t[{\rho(t)}]=-\frac{i}{\hbar}[H_S(t),\rho(t)]\\
&&+\sum^{n^2-1}_{\alpha=1}c_\alpha(t)\left(L_\alpha(t)\rho(t)L_\alpha(t)^\dag-
 \frac{1}{2}\left\{L_\alpha^\dag(t)L_\alpha(t),\rho(t)\right\}\right), \nonumber
\end{eqnarray}
where $n$ is the finite dimension of the open system, 
$H_{S}(t)=H_S^{\dag}(t)$ and $L(t)$ are possibly time-dependent operators on $\mathbbm{C}^n$, 
and $c_\alpha(t)$ are real functions of time.

For now we restrict to P-divisible evolutions 
\cite{Vacchini2011,Chruscinski2014,Wissmann2015,Bernardes2015},
i.e., the dynamical maps $\Lambda_t=\text{T exp}[{\int_0^t\mathcal{L}_s ds)}]$ ($\text{T}$ is the time ordering operator) can be decomposed as $\Lambda_t = \Phi_{t,s}\circ \Lambda_s$
where $\Phi_{t,s}$ is positive (P), for any $t\geq s$.
Let us stress that
P-divisibility is a weaker requirement than $c_\alpha(t)\geq 0$ for any $\alpha$,
which is precisely the condition guaranteeing that MCWF can be applied.
In fact, the positivity of the coefficients coincides, under some regularity conditions, with the property of
completely positive(CP)-divisibility, i.e.,
that $\Phi_{t,s}$ in the decomposition above is CP \cite{Rivas2010,Laine2010}.
The map $\Phi_{t,s}$ is CP
when $(\Phi_{t,s}\otimes \mathbbm{1}_{n}) \rho_{sa} \geqslant 0$, where $\mathbbm{1}_{n}$ is the identity map 
on the ancillary Hilbert space $\mathbbm{C}^n$ and 
$\rho_{sa}$ is any combined open system and ancilla state \cite{Breuer2002}.
The basic observation, which we need to define the rate operator quantum jump unravelling, is 
that the evolution is P-divisible
if and only if
the rate operator
\begin{align}\label{eq:trmain}
&W_{\psi(t)}^J= \\
&\sum_{\alpha=1}^{n^2-1} c_\alpha(t)(L_\alpha(t)
-\ell_{\psi(t),\alpha})\proj{\psi(t)}(L_\alpha(t)-\ell_{\psi(t),\alpha})^\dag, \nonumber
\end{align}
where $\ell_{\psi(t),\alpha}=\bra{\psi(t)}L_\alpha(t)\ket{\psi(t)}$,
is a positive semi-definite operator for any fixed $\ket{\psi(t)}$ \cite{Caiaffa2017}. 
Then, the eigenvalues of $W^J$ are non-negative and we
define the jump operators
\begin{equation}\label{eq:vmain}
V_{\psi(t),j}=\sqrt{\lambda_{\psi(t),j}}\projj{\varphi_{\psi(t),j}}{\psi(t)},
\end{equation}
with $\lambda_{\psi(t),j}$ and $\ket{\varphi_{\psi(t),j}}$ eigenvalues and (orthonormal) eigenvectors of 
$W_{\psi(t)}^J$.

Now, consider the trajectories on the set of the open system pure states,
which are given by the deterministic evolution
fixed by the non-Hermitian and nonlinear Hamiltonian 
\begin{align}\label{eq:heffmain}
H_{\psi(t)}&=H_S(t)
-\frac{i\hbar}{2}\sum_{\alpha=1}^{n^2-1}c_\alpha(t) \\
& \times
\left(L_\alpha^\dag(t) L_\alpha(t)-2\ell_{\psi(t),\alpha}^*L_\alpha(t)
+|\ell_{\psi(t),\alpha}|^2\right) \nonumber
\end{align}
according to 
\begin{equation}
\ket{\psi(t)} \mapsto \ket{\psi(t+d t)}=\frac{(1-\frac{i}{\hbar} H_{\psi(t)}\mathd t)\ket{\psi(t)}}{\|(1-\frac{i}{\hbar} H_{\psi(t)}\mathd t)\ket{\psi(t)}\|},\label{eq:detmain}
\end{equation}
interrupted by sudden jumps in the form
\begin{equation}\label{eq:jumpsmain}
\ket{\psi(t)}\rightarrow \frac{V_{\psi(t),j}\ket{\psi(t)}}{\|V_{\psi(t),j}\ket{\psi(t)}\|} = \ket{\varphi_{\psi(t),j}},
\end{equation}
where the probability to have a jump $j$ between $t$ and $t+\mathd t$ is
\begin{equation}\label{eq:probmain}
p_j(t)=\|V_{\psi(t),j}\ket{\psi(t)}\|^2\mathd t = \lambda_{\psi(t),j} \mathd t.
\end{equation}
As shown in Appendix \ref{app:ufp}, 
this defines a legitimate unravelling, 
i.e., the state
averaged over the different trajectories
satisfies the master equation (\ref{eq:me}).

Such construction resembles the standard MCWF.
But now, crucially,
the different jump operators and their occurrence probabilities are fixed by the
eigenvectors $\ket{\varphi_{\psi(t),j}}$
and eigenvalues $\lambda_{\psi(t),j}$ of the operator $W_{\psi(t)}^J$, rather than by the operators $L_\alpha(t)$
and coefficients $c_\alpha(t)$ as in MCWF.
This is why we can have positive probabilities in Eq.(\ref{eq:probmain}) also for some dynamics
with at least one negative rate $c_\alpha(t)$, where MCWF cannot be applied.
Let us stress that jump-like unravellings for P semigroups (i.e. under the further assumption
that $\Phi_{t,s}=\Lambda_{t-s}$)
have been introduced in \cite{Diosi1986,Diosi1988}, while diffusive unravellings
were defined in \cite{Diosi1988,Gisin1990} and, 
for the more general case of P-divisible dynamics, in \cite{Caiaffa2017}.


\textit{Continuous-measurement interpretation.---}%
To introduce a 
proper continuous-measurement interpretation \cite{Barchielli1991}, let us consider the following setup.
The open system of interest is surrounded by $n$ measurement
apparata, say $n$ counters, which monitor it
continuously and are parametrized by the index $j$.
In the current case, the $n$ apparata correspond to the eigenstates of the rate operator in Eq.~\eqref{eq:trmain}. If a given detector
``clicks'', this means that the state of the system jumps to the corresponding eigenstate, i.e., the detectors count the jumps to the eigenstates of the rate operator.
In the case of no detection at a given moment of time, the evolution continues deterministically.

The type and instant of the counts up to time $t$
define different sequences $\omega_t  = (t_1, j_1; t_2, j_2; \ldots t_m, j_m)$,
with $t_1\leq \ldots \leq t_m \leq t$.
So let $\mathcal{O}=\left\{{\small{\emptyset}}, j\right\}_{j=1, \ldots, n}$ be
the set of measurement outcomes,
where $j$ indicates that the counter $j$ clicked,
while $\emptyset$ that no counter clicked. 
For any time $t$ and sequence $\omega_t$,
we define the
instrument \cite{Heinosaari2012} which maps any element of
$\mathcal{O}$
to an open-system operation, i.e., CP trace non-increasing map,
$\left\{\mathcal{I}_{\omega_t,\footnotesize{\emptyset}},\mathcal{I}_{\omega_t,j}\right\}_{j=1, \ldots, n}$.
The latter fixes the state transformation 
$\rho \mapsto \mathcal{I}_{\omega_t,j (\footnotesize{\emptyset})} \rho/\mbox{Tr}\left\{ \mathcal{I}_{\omega_t,j(\footnotesize{\emptyset})} \rho\right\}$
and probability
$p_{j (\footnotesize{\emptyset})}(t)= \mbox{Tr}\left\{ \mathcal{I}_{\omega_t,j(\footnotesize{\emptyset})} \rho\right\}$
associated with the outcome $j$ ($\emptyset$);
we restrict to purity-preserving transformations.  
As a result 
of the continuous measurement,
the open system, initially in a pure state $\ket{\psi(t_0)}$,
will follow the evolution $\ket{\psi(\omega_t)}$ obtained
by applying every infinitesimal time $\mathd t$ one of the operations 
in $\left\{\mathcal{I}_{\omega_t,\footnotesize{\emptyset}},\mathcal{I}_{\omega_t,j}\right\}_{j=1, \ldots, n}$
(and normalizing the resulting state), 
according to the count sequence $\omega_t$.

In particular, for any time $t$ and sequence $\omega_t$, we define the operation
associated to the count $j$
between $t$ and $t+\mathd t$ as
 \begin{equation}\label{eq:inst}
 \mathcal{I}_{\omega_t,j} \rho = V_{\omega_t, j} \rho V^{\dag}_{\omega_t, j} \mathd t \qquad j=1, \ldots n,
 \end{equation}
which is indeed CP and trace non-increasing; here $V_{\omega_t, j}$
is a short-hand notation for $V_{\psi(\omega_t), j}$, which is defined as in \reff{eq:vmain},
with respect to the state $\ket{\psi(\omega_t)}$ (see also Appendix \ref{app:cmi}).
Moreover, let
\begin{equation}\label{eq:null}
 \mathcal{I}_{\omega_t,\footnotesize{\emptyset}} \rho = F_{\omega_t,\footnotesize{\emptyset}} \rho F^{\dag}_{\omega_t,\footnotesize{\emptyset}}
\end{equation}
be the operation associated with the ``null-count''.
As a defining property of any instrument, 
the overall probability has to be 1.
By virtue of Eqs.(\ref{eq:trmain})-(\ref{eq:heffmain}), 
one can see that this is achieved  by
defining
\begin{equation}\label{eq:fet}
 F_{\omega_t,\footnotesize{\emptyset}} =\left(\mathbbm{1}-\frac{i}{\hbar}H_{\omega_t} \mathd t\right) \Pi_{\omega_t},
\end{equation}
where $\Pi_{\omega_t}=\proj{\psi(\omega_t)}$,
and introducing an auxiliary event $a$,
associated with
$\mathcal{I}_{\omega_t,\footnotesize{a}} \rho = (\mathbbm{1}-\Pi_{\omega_t}) \rho(\mathbbm{1}-\Pi_{\omega_t})$, so that 
$\sum_{j=1}^{n} p_j(t)+p_{\footnotesize{\emptyset}}(t)+p_{\footnotesize{a}}(t)=1$ for any $\rho$.

Now, when applied to the pure state $\rho=\proj{\psi(\omega_t)}$,
the state transformation and occurrence probability fixed by 
Eq.(\ref{eq:inst}) coincide with, respectively, Eq.(\ref{eq:jumpsmain}) and Eq.(\ref{eq:probmain}),
while the state transformation due to \reffs{eq:null} and (\ref{eq:fet}) coincides with
the deterministic one in \reff{eq:detmain};
indeed, the auxiliary event $a$ happens with probability 0.
We can thus conclude that, for any sequence of counts $\omega_t$, the open-system state obtained
by applying $\mathcal{I}_{\omega_t,j}$ and $\mathcal{I}_{\omega_t,\footnotesize{\emptyset}}$ 
every infinitesimal time $\mathd t$
and resulting in $\ket{\psi(\omega_t)}$
provides us with the same
trajectories and associated probabilities as the unravelling described in the previous paragraph
(identifying $\omega_t$ with the sequence of jumps).
In Appendix \ref{app:cmi}
we also give the description
of the above continuous-measurement
evolution in terms of a stochastic differential equation \cite{Barchielli1991}.

Let us stress that 
Eqs.(\ref{eq:inst}) and (\ref{eq:null}) define a family of instruments, one for every time $t$ and sequence
$\omega_t$.
In the standard approaches \cite{Barchielli1991,Dalibard1992,Plenio1998} the probabilities to have a certain
count $j$ given the sequence $\omega_t$ do depend on $\omega_t$, i.e., 
they are to be understood as conditional probabilities \cite{Barchielli1991};
in rate operator jumps, in addition to this, the instrument itself, and then the resulting
state transformation, becomes an object conditioned on $\omega_t$,
in principle different for any count sequence.
This is the key feature which allows us to introduce consistently and systematically a measurement
interpretation for a class of dynamics, the P-divisible ones, 
which is strictly larger than the set of CP-divisible dynamics, where the standard scheme applies.
Later on, we will discuss the meaning of this dependence of the instrument on $\omega_t$ 
in terms of memory effects in the unravelling.

\textit{General open quantum system dynamics: reverse quantum jumps.---}
We now move on to the second main purpose of the paper, that is,
introducing a general version of the rate operator quantum jump method, able to deal also with non-P-divisible dynamics.

When P-divisibility is broken, the rate operator $W^J$ in Eq.(\ref{eq:trmain})
is not positive definite, but it is still Hermitian and
we can thus write its spectral decomposition 
as:
\begin{align}
W_{\psi(t)}^J
&=\sum_{j^+}\lambda_{\psi(t),j^+}\proj{\varphi_{\psi(t),j^+}}\nonumber\\
&-\sum_{j^-}\left|\lambda_{\psi(t),j^-}\right|\proj{\varphi_{\psi(t),j^-}},\label{eq:psng2main}
\end{align}
where $\lambda_{\psi(t),j^+}$ ($\ket{\varphi_{\psi(t),j^+}}$)
and $\lambda_{\psi(t),j^-}$ ($\ket{\varphi_{\psi(t),j^-}}$) are the positive and negative eigenvalues (eigenvectors)
of $W_{\psi(t)}^J$, respectively.
Once again, we define the rate operator jump unravelling as the deterministic evolution
fixed by Eqs.(\ref{eq:heffmain}) and (\ref{eq:detmain}) interrupted by sudden jumps,
associated to the spectral decomposition of $W^J$.
For the positive eigenvalues $\lambda_{\psi(t),j^+}$, we can proceed exactly as
in the P-divisible case, introducing the operators
$V_{\psi(t),j^+}$ as in \reff{eq:vmain}, which induce the jump in \reff{eq:jumpsmain}
with probability as in \reff{eq:probmain}.
On the other hand, for the negative eigenvalues
$\lambda_{\psi(t),j^-}$ we cannot proceed in the same way, as we would get negative probabilities
(analogously to what happens in MCWF for negative coefficients in the master equation).
A possible way out is obtained by relating
the different trajectories of the unravelling to each other \cite{Piilo2008}.
Hence, let us consider the ensemble $\Psi(t)=\left\{\ket{\psi_i(t)}\right\}_{i=1,\ldots,N}$
of the pure states generated by the $N$ trajectories
of the unravelling at time $t$. 
We define a second kind of jump operator, given by
\begin{align}
B_{\psi_k(t),\psi_{k'}(t),{j^-}}&=\sqrt{\left|\lambda_{\psi_{k'}(t),j^-}\right|}\projj{\psi_{k'}(t)}{\psi_{k}(t)},
\label{eq:vtmain}
\end{align}
and we postulate that it acts only if 
the source and target states are related by
\begin{equation}\label{eq:cond}
\ket{\psi_{k}(t)} = \ket{\varphi_{\psi_{k'}(t),j^-}},
\end{equation} inducing the state transformation
$\ket{\psi_{k}(t)}\mapsto \ket{\psi_{k'}(t)}$,
with probability
\begin{align}
p_{j^-}^{(k\rightarrow{k'})}(t)
&=\frac{N_{k'}(t)}{N_k(t)}\left|\lambda_{\psi_{k'}(t),j^-}\right| \mathd t\label{eq:probmmain},
\end{align}
where $N_i(t)$ is the number of elements $\ket{\psi_i(t)}$ in $\Psi(t)$.
In Appendix \ref{app:gm},
we show that the
trajectories described above do provide a valid unravelling, i.e.,
the average state $\sum_i N_i(t)\proj{\psi_i(t)}/N$ satisfies the master equation~(\ref{eq:me}).

Differently from the jumps in \reff{eq:vmain}, each of the jumps in \reff{eq:vtmain} connects
couples of states 
($\ket{\psi_k(t)}$ and $\ket{\psi_{k'}(t)}$) which must be both in
the ensemble $\Psi(t)$ before the jump, and
the associated probability depends on the number of corresponding
ensemble members ($N_k(t)$ and $N_{k'}(t)$), see \reff{eq:probmmain}.
The crucial point is that only if the source state $\ket{\psi_{k}(t)}$
is related to the target state $\ket{\psi_{k'}(t)}$ by the relation in \reff{eq:cond}
they will be connected by a jump $B_{\psi_k(t),\psi_{k'}(t),{j^-}}$.
Note that this also means that this kind of jump can be interpreted as a reverse jump, with respect to
 the ``standard'' ones.
The extension of the rate operator quantum jump method to non-P-divisible dynamics
is in fact inspired by the reverse quantum jumps of
the NMQJ method \cite{Piilo2008,Piilo2009}, but, as will be shown explicitly below, 
ROQJ has a wider range of applicability.

\begin{center} 
\begin{figure}[t!]\label{dephasing}
\includegraphics[width=0.5\textwidth]{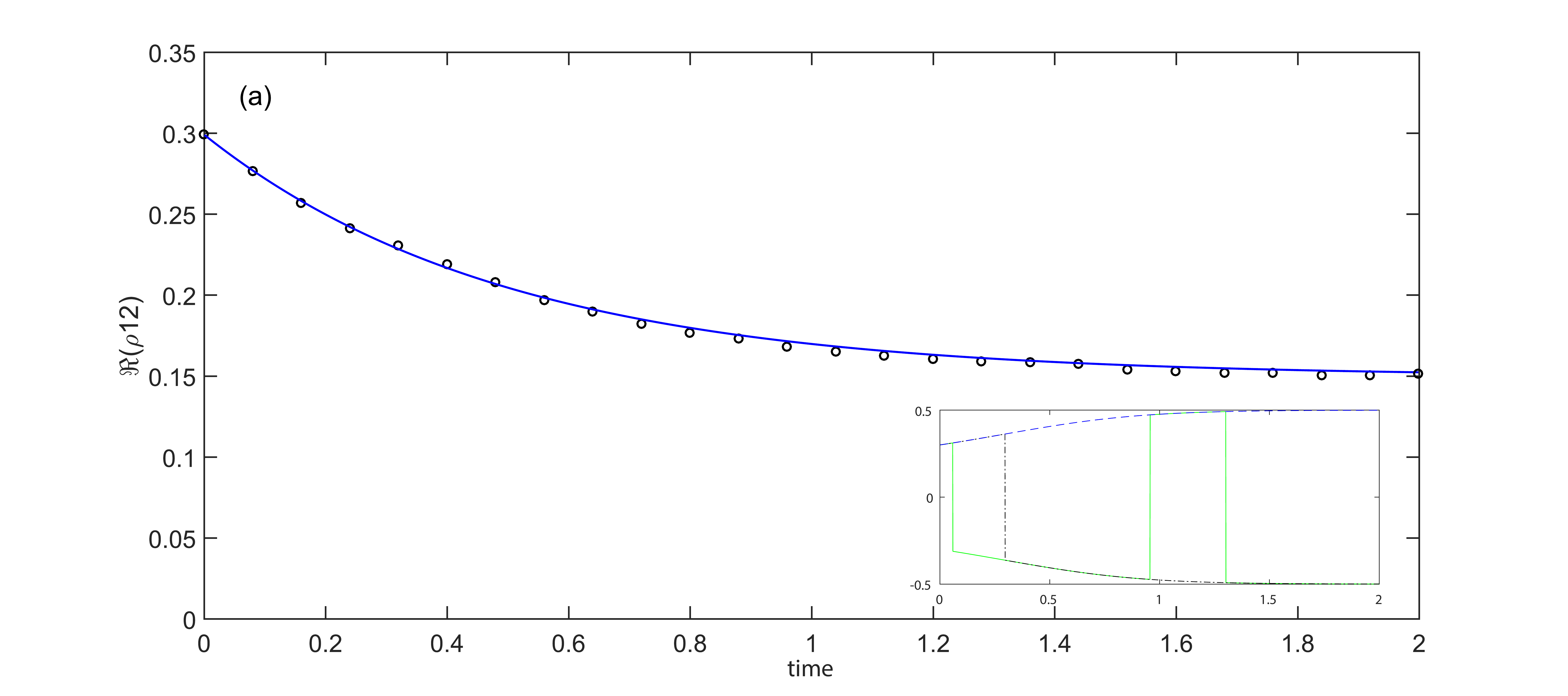}
\\ \includegraphics[width=0.5\textwidth]{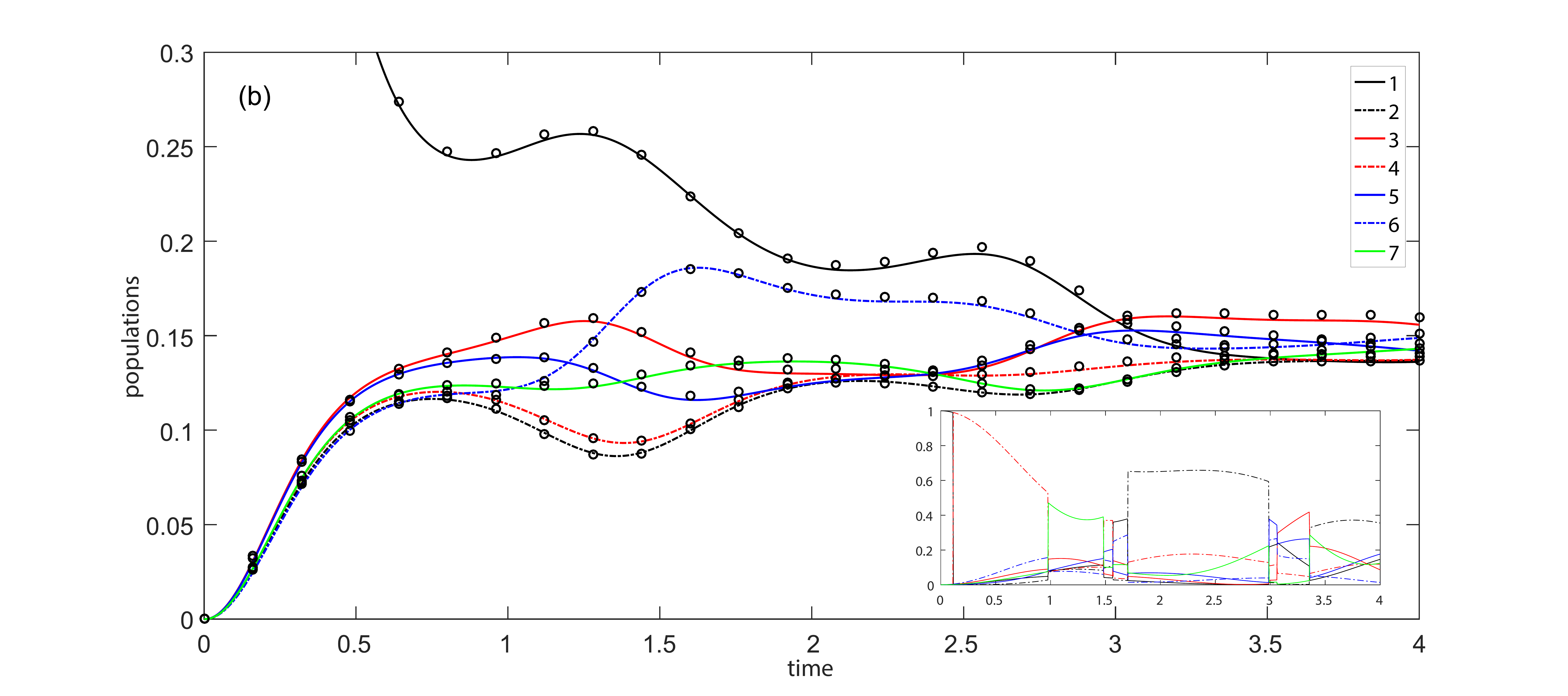}
\caption{(color online) (a) Evolution of the real part of the 2-level system coherence, $\Re\left(\rho_{12}(t)\right)$, according to the master equation in Eq.(\ref{eq:meex}) (solid line), and as average of $10^4$ trajectories (circels) with $dt=0.002$.
The decay rates are $\gamma_1(t)=\gamma_2(t)=1$, and $\gamma_3=-\tanh (t)<0$ for $t>0$. Inset: Three examples of realizations: Evolution 
of $\Re\left(\alpha(t)\beta^*(t)\right)$ with $\ket{\psi(t)}=\alpha(t)\ket{1}+\beta(t)\ket{0}$.
(b) Dissipative 7-coupled-site system (for more details see the text). The simulation results for the site populations (circles) show excellent match with the analytical results (solid lines). The system is initially in the pure state $|\psi\rangle = |1\rangle.$ 
We have used ensemble size $3\times 10^4$ and time-step size $dt=0.005$. Inset: An example realization: Evolution of the site populations.
In both cases, the error bars of the simulation results are smaller than the circles.
} \label{fig:1}
\end{figure}
\end{center}

\textit{Two case studies.---}
First, we consider a master equation where one of the decay rates is negative for all times $t>0$ -- while the corresponding dynamical $\Lambda_t$ is CP, not CP-divisible, and still P-divisible for all $t>0$.
In this case, one cannot use MCWF nor NMQJ methods.
Take the dynamics of a 2-level system fixed 
by the master equation \cite{Hall2014,Andersson2007,Chruscinski2015,Megier2017}
\begin{equation}\label{eq:meex}
\frac{\mathd}{\mathd t} \rho(t) =\frac{1}{2}\sum_{k=1}^3\gamma_k(t)\left[\sigma_k\rho(t)\sigma_k-\rho(t)\right],
\end{equation}
where the $\left\{\sigma_k\right\}_{k=1,2,3}$ are the Pauli matrices.
Eq.(\ref{eq:meex}) is exactly solvable and
the P-divisibility of the corresponding evolution is equivalent 
to the conditions \cite{Chruscinski2015,Megier2017} $\gamma_i(t)+\gamma_j(t)\geq0$, $i\neq j$;
the controlled transition between P-divisibile and non-P divisible
evolutions for the dynamics in Eq.(\ref{eq:meex}) has been realized experimentally in \cite{Bernardes2015}.

Let us fix in particular
$\gamma_i(t)=\mu_i(t)-\mu_j(t)-\mu_k(t)$, 
with $\mu_i(t) = -(x_j+x_k)/(x_j+x_k+e^{2t} x_i)$, for $i\neq j \neq k =1,2,3$ and
with $x_1, x_2, x_3$ non-negative numbers summing up to 1.
P-divisibility holds at any time, but 
the rates $\gamma_i(t)$ can be negative; even more, there
are choices of the $x_i$s such that one of the $\gamma_i(t)$ is negative
for any time $t>0$, i.e., CP-divisibility is broken at any $t>0$. Dynamics
with a perpetually negative master-equation coefficient have been extensively studied
in the literature
\cite{Cresser2010,Hall2014,Bernardes2015,Megier2017,Ferialdi2017,Mitchison2018,Torre2018} and are usually referred to as eternal non-Markovian.
This kind of master equations
cannot be unravelled by the standard MCWF~\cite{Dalibard1992} since 
a negative decay rate leads to a negative quantum jump probability.
NMQJ~\cite{Piilo2008,Piilo2009}, in turn, is based on cancelling previously occurred quantum jumps when the decay rate turns negative. Having a negative rate since the very beginning of the time evolution implies 
that one should cancel something that never happened -- which 
leads to the mathematical problems in addressing such jumps.
On the other hand, rate operator quantum jumps
can easily treat such a situation, as shown in Fig.~\ref{fig:1}. 
By choosing  $x_1=x_2=1/2$ and $x_3=0$, the corresponding decay rates in master equation~\eqref{eq:meex} are
 $\gamma_1(t)=1$,  $\gamma_2(t)=1$, and $\gamma_3=-\tanh (t)<0$ for $t>0$.
We report in Fig.~\ref{fig:1} (a) the evolution
of the 2-level system population averaged over $N=10^3$ realizations.
The excellent agreement with the exact solution can be seen on the whole time scale. 
Moreover, some illustrative trajectories are reported in the inset of Fig.~\ref{fig:1}. 
Indeed, the jumps can be read as the action
of the operations defined in Eq.(\ref{eq:inst}),
associated with the ``click'' of a detector which is continuously monitoring the 2-level system.

As second example, we consider a 7-site system including Hamiltonian interaction between the sites and also dissipative jumps between them. The open system Hamiltonian is $H_S=\sum_{i\neq j} \Omega_{i,j} 
|i\rangle \langle j|$, where  $\Omega_{i,j} = \Omega_{j,i}$, and the values for each are chosen uniformly random so that
$0\leqslant \Omega_{i,j} \leqslant 0.6$. In other words, all of the sites are coupled unitarily to all other sites with random coupling strength. In the dissipator,  jumps can happen between any pair of sites $i$ and $j$, i.e.,
we have 49 jump operators given by $c(t)  |i\rangle \langle j|$ for any combination of $i$ and $j$. Here,
for simplicity, we use for all operators equal rate, which we choose as $c(t)=0.5[(1-e^{-0.5t})0.3  +  e^{-0.3t}\sin(4.5t)]$. This also guarantees CP of the dynamical map since the time integral of the rate remains positive. The rate oscillates between positive and negative values, and P-divisibility is broken whenever the rate is negative. By exploiting the properties of the ROQJ method, we have in the simulation only $\sqrt{49}=7$ decay channels. Fig.~\ref{fig:1}(b) shows the excellent match between the analytical and simulation results while the inset displays an example realization.

\textit{In-between Markovian and non-Markovian.---}
We now clarify the different degrees of
memory effects present in rate operator quantum jumps, 
also in comparison with other (jump) unravelling approaches.
Let us start from CP-divisible dynamics, which have been identified with quantum
Markovian dynamics in \cite{Rivas2010}.
Here, MCWF can be applied and the resulting unravelling
is build up via the same non-Hermitian Hamiltonian and Lindblad
operators
for any sequence of jumps $\omega_t$. 
On the other hand, 
the probability to have a jump $j$ at a time $t$
depends on the state before the jump, $\ket{\psi(\omega_t)}$,
and then on all the previous sequence of types and instants of jumps which led to that state.
We conclude that the jump probabilities do carry some memory~\cite{foot},
though the averaged state can follow, e.g., semigroup dynamics.
  
If we now move to the P-divisible case and the unravelling provided by ROQJ method, 
we see that the memory described above gets amplified, since now not only the probabilities, but also the kind of jump
at a given time depend on the previous sequence of jumps.
In terms of the measurement interpretation, this means that not only the outcome at time $t$, 
but also the measurement
apparatus used to realize a certain instrument will have to depend on the past outcomes.  

The strongest form of memory for the unravellings 
is certainly the one characterizing the reversed jumps, both
in NMQJ and in the non-P-divisible version of ROQJ method.
Here, the jump probabilities and operators connect different trajectories,
in a way that the event at a time $t$ on a given trajectory
will depend on the previous events also on all the other trajectories. No measurement
interpretation is possible in this case.

\textit{Conclusions.---}
In this paper, we have introduced a quantum-jump unravelling, named rate operator quantum jumps, 
which allowed us to define a consistent measurement interpretation 
for a wider class of dynamics than those where the standard MCWF interpretation applies.
This includes the case where the master equation contains negative rates and the corresponding dynamical map is not CP-divisible.
Our approach is able to deal with any open quantum system dynamics -- including
dynamical regions where neither MCWF nor NMQJ can be used -- and provides a unified framework for using quantum jumps to deal with open system dynamics. 
Moreover, our results highlight the different kinds of memory effects which arise
within the context of quantum unravellings and will hopefully help further clarify 
the difference between Markovianity and non-Markovianity in the quantum realm.

\textit{Acknowledgments}
This work was supported by the FFABR project of MIUR,
the ERC Synergy grant BioQ and the Academy of Finland via the Centre of Excellence program 
(Project no. 312058 as well as Project no. 287750).
The authors would like to thank Susana Huelga and Martin Plenio for useful discussions.

\newpage
\appendix

\onecolumngrid
\section{Proof that ROQJ is a proper unravelling of the master equation --- P-divisible case}\label{app:ufp}

P-divisibility means that the eigenvalues $\lambda_{\psi(t),j}$ of the rate operator $W_{\psi(t)}^J$ are non-negative,
so that we can write
\begin{align}
W_{\psi(t)}^J&=\sum_{j=1}^{n}\lambda_{\psi(t),j}\proj{\varphi_{\psi(t),j}} \nonumber\\
&=\sum_{j=1}^{n}V_{\psi(t),j}\proj{\psi(t)}V_{\psi(t),j}^\dag,\label{eq:sd}
\end{align}
where we have defined [see also Eq.(\ref{eq:vmain}) of the main text]
\begin{equation}\label{eq:v}
V_{\psi(t),j}=\sqrt{\lambda_{\psi(t),j}}\projj{\varphi_{\psi(t),j}}{\psi(t)},
\end{equation}
which describes a jump from the current state $\ket{\psi(t)}$
to one of the orthogonal eigenvectors of $W_{\psi(t)}^J$, $\ket{\varphi_{\psi(t),j}}$;
note that, indeed, both the eigenvalues and the eigenvectors of $W_{\psi(t)}^J$
will generally depend on the state $\ket{\psi(t)}$;
moreover, it is easy to see that $\ket{\psi(t)}$ itself is an eigenvector of $W_{\psi(t)}^J$,
with respect to the eigenvalue 0.

Given a deterministic evolution governed by the non-Hermitian Hamiltonian 
[see Eq.(\ref{eq:heffmain}) of the main text]
\begin{equation}\label{eq:heff}
H_{\psi(t)}=H_S-\frac{i\hbar}{2}\sum_{\alpha=1}^{n^2-1}c_\alpha(t)\left(L_\alpha^\dag(t) L_\alpha(t)-2\ell_{\psi(t),\alpha}^*L_{\alpha}(t)+
|\ell_{\psi(t),\alpha}|^2\right),
\end{equation}
for a small time step $\mathd t$, $\ket{\psi(t)}$ evolves according to
\begin{align}
\ket{\psi(t+\mathd t)}&=\frac{\ket{\phi(t+\mathd t)}}{\|\ket{\phi(t+\mathd t)}\|},\label{eq:dete}\\
\text{where\ \ \ }\ket{\phi(t+\mathd t)}&=\left(1-\frac{iH_{\psi(t)}\mathd t}{\hbar}\right)\ket{\psi(t)}\nonumber\\
&=\left[1-\frac{iH_S\mathd t}{\hbar}-\frac{\mathd t}{2}\sum_{\alpha=1}^{n^2-1}c_\alpha(t)\left(L_\alpha^\dag(t) L_\alpha(t)-2\ell_{\psi(t),\alpha}^*L_\alpha(t)+
|\ell_{\psi(t),\alpha}|^2\right)\right]\ket{\psi(t)}\label{eq:det}.
\end{align}
As defined in the main text, in ROQJ the evolution above is interrupted by sudden jumps of the form 
[see Eq.(\ref{eq:jumpsmain}) of the main text]
\begin{equation}\label{eq:jumps}
\ket{\psi(t)}\rightarrow \frac{V_{\psi(t),j}\ket{\psi(t)}}{\|V_{\psi(t),j}\ket{\psi(t)}\|}
\end{equation}
which happen with probability [see Eq.(\ref{eq:probmain}) of the main text]
\begin{equation}\label{eq:prob}
p_j(t)=\|V_{\psi(t),j}\ket{\psi(t)}\|^2\mathd t.
\end{equation}
It also follows that the deterministic evolution must occurs with probability $1-P_{\text{jump}}(t)$ instead, where $P_{\text{jump}}(t)=\sum_{j=1}^{n}p_j(t)$. Moreover, notice the important relation 
\begin{equation}\label{eq:norm}
1-P_{\text{jump}}(t)=\|\ket{\phi(t+\mathd t)}\|^2,
\end{equation}
where $\ket{\phi(t+\mathd t)}$ is the unnormalized state of \reff{eq:det}.

To show that ROQJ provides us with a legitimate unravelling of the master equation, 
we shall consider the value of the state of the system averaged over the different 
trajectories of the piecewise deterministic process described above, weighted by their occurrence probability. 
Note that if we start from a pure state $\ket{\psi(0)}$, since both the deterministic and the jump part of the
evolution preserve the purity, we will have a pure state on any single trajectory at any time;
indeed, the average state will instead be mixed. 
It is convenient to perform the average in two steps.
First, we fix the state $ \ket{\psi(t)}$ at time $t$
and we perform the average (denoted as $\overline{\cdot}$) 
of the state $\ket{\xi(t+\mathd t)}$ which we have at time $t+\mathd t$,
conditioned on having $ \ket{\psi(t)}$ at time $t$, thus getting
\begin{equation}\label{eq:avg}
\overline{\proj{\xi(t+dt)}}=\left(1-\sum_{j=1}^{n}p_j(t)\right)\proj{\psi(t+\mathd t)}+\sum_{j=1}^{n}p_j(t)\frac{V_{\psi(t),j}\proj{\psi(t)}V_{\psi(t),j}^\dag}{\|V_{\psi(t),j}\ket{\psi(t)}\|^2}:
\end{equation}
$\ket{\xi(t+dt)}$ can be identified with $\ket{\psi(t+\mathd t)}$ if the deterministic
evolution occurs, which happens with probability $1-\sum_{j=1}^{n}p_j(t)$,  
and with $V_{\psi(t),j}\ket{\psi(t)}/\|V_{\psi(t),j}\ket{\psi(t)}\|$ if the jump $j$ occurs,
which happens with probability $p_j(t)$.
For what the deterministic part is concerned, using Eqs.~\eqref{eq:det} and \eqref{eq:norm} and omitting the terms in $\mathd t^2$ we get 
\begin{align}
&\left(1-\sum_{j=1}^{n}p_j(t)\right)\proj{\psi(t+\mathd t)}=\left(1-P_{\text{jump}}(t)\right)\frac{\proj{\phi(t+\mathd t)}}{\|\phi(t+\mathd t)\|^2}\nonumber\\
&=\left(1-\frac{iH_{\psi(t)}\mathd t}{\hbar}\right)\proj{\psi(t)}\left(1-\frac{iH_{\psi(t)}\mathd t}{\hbar}\right)^\dag\nonumber\\
&=\proj{\psi(t)} - 
\frac{i}{\hbar}[H_S,\proj{\psi(t)}]\mathd t-\frac{1}{2}\sum_{\alpha=1}^{n^2-1}c_\alpha(t)
\biggl(\left\{L_\alpha^\dag(t)L_\alpha(t),\proj{\psi(t)}\right\} 
-2\ell_{\psi(t),\alpha}^*L_\alpha(t)\proj{\psi}
\nonumber\\
& -2\ell_{\psi(t),\alpha}\proj{\psi}L_\alpha^\dag(t)+2|\ell_{\psi(t),\alpha}|^2\proj{\psi}|
\biggr)
\mathd t\label{eq:det_term}.
\end{align}
On the other hand, the jump term of \reff{eq:avg} reads 
\begin{align}
\sum_{j=1}^{n}p_j(t)\frac{V_{\psi(t),j}\proj{\psi(t)}V_{\psi(t),j}^\dag}{\|V_{\psi(t),j}\ket{\psi(t)}\|^2}&
=\sum_{j=1}^{n}\|V_{\psi(t),j}\ket{\psi(t)}\|^2\frac{V_{\psi(t),j}\proj{\psi(t)}V_{\psi(t),j}^\dag}{\|V_{\psi(t),j}\ket{\psi(t)}\|^2}\mathd t\nonumber\\
&=\sum_{j=1}^{n}V_{\psi(t),j}\proj{\psi(t)}V_{\psi(t),j}^\dag\mathd t\nonumber\\
&=W_{\psi(t)}^J\mathd t\label{eq:jump_term}.
\end{align}
Putting Eqs.~\eqref{eq:det_term} and \eqref{eq:jump_term} together, we have
\begin{align}
\overline{\proj{\xi(t+dt)}}=
\proj{\psi(t)}-
\frac{i}{\hbar}[H_S,\proj{\psi(t)}]\mathd t-\frac{1}{2}\sum_{\alpha=1}^{n^2-1}
c_\alpha(t)
\biggl(\left\{L_\alpha^\dag(t)L_\alpha(t),\proj{\psi(t)}\right\} 
\nonumber\\
-2\ell_{\psi(t),\alpha}^*L_\alpha(t)\proj{\psi}
-2\ell_{\psi(t),\alpha}\proj{\psi}L_\alpha^\dag(t)+2|\ell_{\psi(t),\alpha}|^2\proj{\psi}
\biggr)
+W_{\psi(t)}^J
\mathd t.
\end{align}
Finally, we perform a second average, this time with respect to the possible states $\ket{\psi(t)}$ over which
we conditioned.
At the left hand side of the equation above we thus simply get the state at time $t+\mathd t$
averaged over all the possible trajectories, $\rho(t+\mathd t)$, while at the right hand side we get
$\rho(t)+\mathcal{L}_t[\rho(t)]\mathd t$, so that we recover
exactly the master equation fixed by
Eq.(1) of the main text (together with the fixed initial condition $\rho(0)=\proj{\psi(0)}$;
of course, the unravelling for a mixed initial state $\rho(0)=\sum_i p_i \proj{\psi_i}$ 
can be obtained by averaging the unravellings for each initial pure state $\proj{\psi_i}$
over the probability distribution given by the $p_i$s).


\section{Continuous-measurement interpretation: stochastic differential equation}\label{app:cmi}
We give here some further mathematical details about the continuous-measurement interpretation
of ROQJ described in the main text, especially providing the corresponding stochastic differential equation (SDE);
indeed the reader is also referred to \cite{Barchielli1991}.

Let $\left\{N_j(t)\right\}_{j=1 \dots n-1}$ be a family of independent counting processes defined on a common probability space 
$(\Omega, \mathcal{F}, \mathbbm{P})$ and $\mathbbm{E}[\cdot]$
the statistical mean with respect to the probability $\mathbbm{P}$.
Furthermore, the trajectories of the counting processes up to time $t$ are indicated
as $\omega_t  = (t_1, j_1; t_2, j_2; \ldots t_m, j_m)$, denoting the types and instants of counts
(and thus identifying with the sequences mentioned in the main text).
The counting processes satisfy the following relations:
\begin{eqnarray}
\mathd N_j(t) \mathd t &=&0\nonumber\\
\mathd N_j(t) \mathd N_k(t) &=& \delta_{jk}\mathd N_j(t)\nonumber\\
\mathbbm{E}[\mathd N_j(t) | \omega_t] &=& \|V_{\psi(t),j}\ket{\psi(t)}\|^2\mathd t, \label{eq:edn3}
\end{eqnarray}
where $\mathd N_j(t) = N_j(t + \mathd t) - N_j(t)$ is the (Ito) increment of $N_j(t)$ in a time $\mathd t$
and $\mathbbm{E}[\cdot | \omega_t]$ is the expectation value \emph{conditioned}
on the trajectory up to time $t$, while $\ket{\psi(t)}$ is the state
satisfying the SDE
\begin{eqnarray}
\mathd \ket{\psi(t)} &=& \left[- \frac{i}{\hbar}H_S- \frac{1}{2}\sum_{\alpha=1}^{n^2-1}c_\alpha(t)\left(L_\alpha^\dag(t) L_\alpha(t)
-2\ell_{\psi(t),\alpha}^*L_\alpha(t)-\bra{\psi(t)}L_\alpha^\dag(t) L_\alpha(t)\ket{\psi(t)}+2|\ell_{\psi(t),\alpha}|^2\right)
\right] \ket{\psi(t)} \mathd t \nonumber\\
&&+\sum_{j=1}^{n}\left(\frac{V_{\psi(t),j}}{\|V_{\psi(t),j}\ket{\psi(t)}\|}-\mathbbm{1}\right) \ket{\psi(t)} \mathd N_j(t),
\label{eq:sdeq2}
\end{eqnarray}
where $\ell_{\psi(t),\alpha}=\bra{\psi(t)}L_\alpha(t)\ket{\psi(t)}$.
Such SDE preserves the normalization of the state $\ket{\psi(t)}$
and, most importantly, the latter should now be seen as a function of the trajectory up to time
$t$, so that as a matter of fact $\ket{\psi(t)}$ is a short-hand notation for $\ket{\psi(\omega_t)}$.
It is easy to see that the SDE defined by Eqs.(\ref{eq:edn3}) and (\ref{eq:sdeq2})
is in fact equivalent to the unravelling fixed by Eqs.\eqref{eq:dete}-\eqref{eq:prob}. 
For the deterministic part, one has simply to use that
$$
\frac{1- \kappa_1 \mathd t}{\sqrt{1- \kappa_2 \mathd t +\kappa_3 \mathd t^2}}
= 1+\left(\frac{\kappa_2}{2}-\kappa_1\right) \mathd t + o(\mathd t^2),
$$
while $(\frac{V_{\psi(t),j}}{\|V_{\psi(t),j}\ket{\psi(t)}\|}-\mathbbm{1})\ket{\psi(t)}$
is indeed the state change induced by the jump in \reff{eq:jumps}
and the conditional expectation values in \reff{eq:edn3} precisely correspond to the
event probabilities in \reff{eq:prob}, since the probability
of having more than one count in a time interval $\mathd t$ is of order $\mathd t^2$
\cite{Barchielli1991}.

\section{Proof that ROQJ is a proper unravelling of the master equation --- General case}\label{app:gm}

We now consider the case of a dynamical map $\Lambda_t=\text{T exp}({\int_0^t\mathcal{L}_s ds)})$ 
which needs not be P-divisible.
We aim to simulate the solution of 
the corresponding master equation in Eq.(\ref{eq:me}) by averaging the dynamics of the pure states of the ensemble 
$\Psi(t)=\left\{\ket{\psi_i(t)}\right\}_{i=1,\ldots,N}$,
\begin{equation}\label{eq:ens}
\varrho(t)=\sum_k\frac{N_k(t)}{N}\proj{\psi_k(t)},
\end{equation}
where $N_k(t)$ is the number of elements $\ket{\psi_k(t)}$ in the ensemble
and indeed $N=\sum_k N_k(t)$ at any $t$.
Using the hermiticity of $W^J$, its spectral decomposition 
can be divided in the positive and negative parts [see Eq.(\ref{eq:psng2main}) in the main text]:
\begin{align}
W_{\psi(t)}^J&=\sum_{j=1}^{n}\lambda_{\psi(t),j}\proj{\varphi_{\psi(t),j}}\nonumber\\
&=\sum_{j^+}\lambda_{\psi(t),j^+}\proj{\varphi_{\psi(t),j^+}}
-\sum_{j^-}\left|\lambda_{\psi(t),j^-}\right|\proj{\varphi_{\psi(t),j^-}},\label{eq:psng2}
\end{align}
where $\lambda_{\psi(t),j^+}$ and $\lambda_{\psi(t),j^-}$ are the positive and negative eigenvalues of $W_{\psi(t)}^J$, respectively,
and $\ket{\varphi_{\psi(t),j^+}}, \ket{\varphi_{\psi(t),j^-}}$ the corresponding orthonormal eigenvectors.

The ROQJ unravelling is then composed of three parts.
First,
a deterministic evolution governed by the non-Hermitian Hamiltonian as in \reff{eq:heff},
so that for an infinitesimal time-step $\mathd t$ the element of the ensemble $\ket{\psi_k(t)}$ evolves according to
\begin{align}
\ket{\psi_k(t+\mathd t)}&=\frac{\ket{\phi_k(t+\mathd t)}}{\|\ket{\phi_k(t+\mathd t)}\|},\label{eq:dete2}\\
\text{where\ \ \ }\ket{\phi_k(t+\mathd t)}&=\left(1-\frac{iH_{\psi(t)}\mathd t}{\hbar}\right)\ket{\psi_k(t)}\nonumber\\
&=\left[1-\frac{iH_S\mathd t}{\hbar}-\frac{\mathd t}{2}\sum_{\alpha=1}^{n^2-1}
c_\alpha(t)\left(L_\alpha^\dag(t) L_\alpha(t)-2\ell_{\psi_k(t),\alpha}^*L_\alpha(t)+
|\ell_{\psi_k(t),\alpha}|^2\right)\right]\ket{\psi_k(t)}\label{eq:det2}.
\end{align}
The evolution above is interrupted by sudden jumps which, for positive $\lambda_{j^+}$, 
are fixed by the forward jump operator
\begin{align}
V_{\psi_k(t),{j^+}}&=\sqrt{\lambda_{\psi_k(t),j^+}}\projj{\varphi_{\psi_k(t),j^+}}{\psi_k(t)\psp},\label{eq:v2}
\end{align}
via
\begin{equation}\label{eq:jumpsp}
\ket{\psi_k(t)}\rightarrow \frac{V_{\psi_k(t),j^+}\ket{\psi_k(t)}}{\|V_{\psi_k(t),j^+}\ket{\psi_k(t)}\|}
\end{equation}
and happen with probability 
\begin{align}
p_{j^+}^{(k)}(t)&=\|V_{\psi_k(t),j^+}\ket{\psi_k(t)}\|^2\mathd t\nonumber\\
&=\lambda_{\psi_k(t),j^+}\mathd t;\label{eq:probp}
\end{align}
of course, these jumps coincide with those for the P-divisibile case in Eqs.(\ref{eq:jumps}) and (\ref{eq:prob}).
Instead, 
for those eigenvalues of the rate-operator arising from the lack of P-divisibility of the master equation,
we define the backward jump operators via Eq.(\ref{eq:vtmain}) of the main text, which can also be restated as
\begin{align}
B_{\psi_k(t),\psi_{k'}(t),{j^-}}&=\sqrt{\left|\lambda_{\psi_{k'}(t),j^-}\right|}\projj{\psi_{k'}(t)}{\psi_{k}(t)}
\delta\left(\ket{\psi_{k}(t)} -\ket{\varphi_{\psi_{k'}(t),j^-}}\right)\label{eq:vt}
\end{align}
to emphasize that
the jumps
\begin{equation}\label{eq:jumpsm}
\ket{\psi_{k}(t)}\rightarrow \ket{\psi_{k'}(t)}
\end{equation}
are constrained by the requirement that 
the source state has to be of the form $\ket{\psi_{k}(t)} = \ket{\varphi_{\psi_{k'}(t),j^-}}$.
The related probability is [see Eq.(\ref{eq:probmmain}) of the main text]
\begin{align}
p_{j^-}^{(k\rightarrow{k'})}(t)&=\frac{N_{k'}(t)}{N_k(t)}\|B_{\psi_k(t),\psi_{k'}(t),j^-}\ket{\psi_{k}(t)}\|^2\mathd t\nonumber\\
&=\frac{N_{k'}(t)}{N_k(t)}\left|\lambda_{\psi_{k'}(t),j^-}\right|
\delta\left(\ket{\psi_{k}(t)} -\ket{\varphi_{\psi_{k'}(t),j^-}}\right) \mathd t\label{eq:probm}.
\end{align}

In order to show the equivalence of this approach with the master equation,
as for the P-divisible case, we shall average over the possible
trajectories described above.
However, since now the different trajectories are possibly connected to each other via
the reversed quantum jumps, it is convenient to perform
one single collective average, i.e., to consider
\begin{align}
\varrho(t+\mathd t)&=\sum_k\frac{N_k(t+ \mathd t)}{N}\proj{\xi_k(t+dt)}\nonumber\\
&=\sum_{k}\frac{N_k(t)}{N}\left[\left(1-P_{\text{jump}}^{(k)}(t)\right)\proj{\psi_k(t+\mathd t)}
+\sum_{j^+}p_{j^+}^{(k)}(t)\frac{V_{\psi_k(t),j^+}\proj{\psi_k(t)}V_{\psi_k(t),j^+}^\dag}{\|V_{\psi_k(t),j^+}\ket{\psi_k(t)}\|^2}\right.\nonumber\\
&\left. 
+\sum_{j^-,k'}p_{j^-}^{(k\rightarrow k')}(t)\frac{B_{\psi_k(t),\psi_{k'}(t),j^-}\proj{\psi_k(t)}B^{\dag}_{\psi_k(t),\psi_{k'}(t),j^-}}{\|B_{\psi_k(t),\psi_{k'}(t),j^-}\ket{\psi_k(t)}\|^2}\right],\label{eq:avg2}
\end{align}
where we fix a time $t$ and the elements of the ensemble at that time are denoted as $\ket{\psi_k(t)}$,
and
\begin{itemize}
\item $N_k(t+ \mathd t)/N$ is the probability that the state at time $t+\mathd t$ is $\ket{\xi_k(t+dt)}$;
\item $N_k(t)/N *(1-P_{\text{jump}}^{(k)}(t))$, where 
$$P_{\text{jump}}^{(k)}(t)=\sum_{j^+}p_{j^+}^{(k)}(t)+\sum_{j^-,k'}p_{j^-}^{(k\rightarrow{k'})}(t),$$ 
is the joint probability that
the state at time $t$ is $\ket{\psi_k(t)}$ and that there is a deterministic evolution
$\ket{\psi_k(t)}\mapsto \ket{\xi_k(t+dt)}=\ket{\psi_k(t+\mathd t)}$;
\item $N_k(t)/N * p_{j^+}^{(k)}(t)$ is the joint probability that
the state at time $t$ is $\ket{\psi_k(t)}$ and that there is a forward jump
$\ket{\psi_k(t)}\mapsto \ket{\xi_k(t+dt)}=V_{\psi_k(t),j^+}\ket{\psi_k(t)}/\|V_{\psi_k(t),j^+}\ket{\psi_k(t)}\|$;
\item $N_k(t)/N * p_{j^-}^{(k\rightarrow k')}(t)$ is the joint probability that
the state at time $t$ is $\ket{\psi_k(t)}$ and that there is a reversed jump
$\ket{\psi_k(t)}\mapsto \ket{\xi_{k'}(t+dt)}=B_{\psi_k(t),\psi_{k'}(t),j^-}\ket{\psi_k(t)}/\|B_{\psi_k(t),\psi_{k'}(t),j^+}\ket{\psi_k(t)}\|$;
crucially, now the probabilities referred to any of these jumps
will depend on how many elements of the ensemble 
coincide with the target state $\ket{\psi_{k'}(t)}=\ket{\xi_{k'}(t+\mathd t)}$ and how many with
the pre-jump state $\ket{\psi_k(t)}=\ket{\varphi_{\psi_{k'}(t),j^-}}$ before the jump, see \reff{eq:probm}.
\end{itemize}
For convenience, we treat the terms in \reff{eq:avg2} separately, following the procedure given in \cite{Piilo2009}. 

For what the deterministic part is concerned
\begin{align}
\left(1-P^{(k)}_{\text{jump}}(t)\right)\proj{\psi_k(t+\mathd t)}&=\left(1-\sum_{j^+}p_{j^+}^{(k)}(t)+\sum_{j^-,k'}p_{j^-}^{(k\rightarrow{k'})}(t)\right)\frac{\proj{\phi_k(t+\mathd t)}}{\|\phi_k(t+\mathd t)\|^2},
\end{align}
as
\begin{align}
\proj{\phi_k(t+\mathd t)}&=\proj{\psi_k(t)}-\frac{i}{\hbar}[H_S,\proj{\psi_k(t)}] \mathd t-\frac{1}{2}\sum_{\alpha=1}^{n^2-1}c_\alpha(t)\biggl(\left\{L_\alpha^\dag(t)L_\alpha(t),\proj{\psi_k(t)}\right\}
\nonumber\\
& 
-2\ell_{\psi_k(t)\alpha}^*L_\alpha(t)\proj{\psi_k(t)}
-2\ell_{\psi_k(t),\alpha}\proj{\psi_k(t)}L_\alpha^\dag(t)
+2|\ell_{\psi_k(t),\alpha}|^2\proj{\psi_k(t)}
\biggr)\mathd t,
\end{align}
we have three contributions up to $\mathd t$:
\begin{align}
&\left(1-P^{(k)}_{\text{jump}}(t)\right)\proj{\psi_k(t+\mathd t)} \nonumber\\
&=\proj{\psi_k(t)}-\frac{i}{\hbar}[H_S,\proj{\psi_k(t)}] \mathd t-\frac{1}{2}\sum_{\alpha=1}^{n^2-1}c_\alpha(t)\biggl(\left\{L_\alpha^\dag(t)L_\alpha(t),\proj{\psi_k(t)}\right\}
\nonumber\\
& 
-2\ell_{\psi_k(t),\alpha}^*L_\alpha(t)\proj{\psi_k(t)}
-2\ell_{\psi_k(t),\alpha}\proj{\psi_k(t)}L_\alpha^\dag(t)
+2|\ell_{\psi_k(t),\alpha}|^2\proj{\psi_k(t)}
\biggr)\mathd t \nonumber\\
& -\sum_{j^+}p_{j^+}^{(k)}(t) \proj{\psi_k(t)}- \sum_{j^-,k'}p_{j^-}^{(k\rightarrow{k'})}(t)\proj{\psi_k(t)}\nonumber\\
& +\sum_{\alpha=1}^{n^2-1}c_\alpha(t)\bra{\psi_k(t)}
(L_\alpha(t)-\ell_{\psi_k(t),\alpha})^\dag
(L_\alpha(t)-\ell_{\psi_k(t),\alpha})\ket{\psi_k(t)} \proj{\psi_k(t)}\mathd t \nonumber\\
&=\proj{\psi_k(t)}-\frac{i}{\hbar}[H_S,\proj{\psi_k(t)}] \mathd t-\frac{1}{2}\sum_{\alpha=1}^{n^2-1}c_\alpha(t)\biggl(\left\{L_\alpha^\dag(t)L_\alpha(t),\proj{\psi_k(t)}\right\}
\nonumber\\
& 
-2\ell_{\psi_k(t),\alpha}^*L_\alpha(t)\proj{\psi_k(t)}
-2\ell_{\psi_k(t),\alpha}\proj{\psi_k(t)}L_\alpha^\dag(t)
+2|\ell_{\psi_k(t),\alpha}|^2\proj{\psi_k(t)}
\biggr)\mathd t \nonumber\\
& - \sum_{j^-,k'}p_{j^-}^{(k\rightarrow{k'})}(t)\proj{\psi_k(t)}
-\sum_{j^-}|\lambda_{\psi_k(t),j^{-}}| \proj{\psi_k(t)}\mathd t \label{eq:qqu}
\end{align}
where for the second equality we have used the identity 
\begin{equation}
\sum_{\alpha=1}^{n^2-1}c_\alpha(t)\bra{\psi_k(t)}(L_\alpha(t)-\ell_{\psi_k(t),\alpha})^\dag
(L_\alpha(t)-\ell_{\psi_k(t),\alpha})\ket{\psi_k(t)} =
\sum_{j^+}\lambda_{\psi_k(t),j^{+}}(t) -\sum_{j^-}|\lambda_{\psi_k(t),j^{-}}(t)|
\end{equation}
which directly follows from taking the trace in the spectral decomposition of the rate operator in \reff{eq:psng2}
(recall the definition in Eq.(\ref{eq:trmain}) of the main text), and the definition of $p_{j^+}^{(k)}(t)$ in \reff{eq:probp}.
Note that, as in NMQJ \cite{Piilo2008,Piilo2009},
the differential of the deterministic part generates, besides the commutator and anti-commutator
terms of the master equation, a further contribution related to the negative rates, which will combine with the jump
part, giving the remaining term in the master equation.

For the forward jumps term of \reff{eq:avg2}, we have 
\begin{align}
\sum_{j^+}p_{j^+}^{(k)}(t)\frac{V_{\psi_k(t),j^+}\proj{\psi_k(t)}V_{\psi_k(t),j^+}^\dag}{\|V_{\psi_k(t),j^+}\ket{\psi_k(t)}\|^2}
=\sum_{j^+}\lambda_{\psi_k(t),j^+}\proj{\varphi_{\psi_k(t),j^+}} \mathd t\label{eq:jump_term2},
\end{align}
while, the negative jumps term reduces to
\begin{align}
&\sum_{j^-}p_{j^-}^{(k\rightarrow k')}(t)\frac{B^{\dag}_{\psi_k(t),\psi_{k'}(t),j^-}\proj{\psi_k(t)}B_{\psi_k(t),\psi_{k'}(t),j^-}}{\|B_{\psi_k(t),\psi_{k'}(t),j^-}\ket{\psi_k(t)}\|^2}\nonumber\\
&=\frac{N_{k'}(t)}{N_k(t)}\sum_{j^-}\left|\lambda_{\psi_{k'}(t),j^-}\right|
\delta\left(\ket{\psi_{k}(t)} -\ket{\varphi_{\psi_{k'}(t),{j^-}}}\right)
\mathd t\label{eq:jump_back}.
\end{align}

All in all, by putting Eqs.~\eqref{eq:qqu}, \eqref{eq:jump_term2} and \eqref{eq:jump_back} together, 
we get
\begin{align}\label{eq:ell}
\varrho(t+\mathd t)&=\varrho(t)-\frac{i}{\hbar}[H_S,\varrho(t)]\mathd t
\nonumber\\
&-\frac{1}{2}\sum_{\alpha=1}^{n^2-1}c_\alpha(t)
\biggl(\left\{L_\alpha^\dag(t)L_\alpha(t),\varrho(t) \right\}
-2\ell_{\psi_k(t),\alpha}^*L_{\alpha}\varrho(t)
-2\ell_{\psi_k(t),\alpha}\varrho(t)L_{\alpha}^\dag
+2|\ell_{\psi_k(t),\alpha}|^2\varrho(t)
\biggr)\mathd t
\nonumber\\
&+\sum_{k,j^+}\frac{N_k(t)}{N}
\lambda_{\psi_k(t),j^+}\proj{\varphi_{\psi_k(t),j^+}} \mathd t
-\sum_{k,j^-}\frac{N_{k}(t)}{N}|\lambda_{\psi_k(t),j^{-}}| \proj{\psi_k(t)}\mathd t \nonumber\\
&+\sum_{k,k',j^-}\frac{N_{k'}(t)}{N} \left|\lambda_{\psi_{k'}(t),j^-}\right|\proj{\psi_{k'}(t)}
\delta\left(\ket{\psi_{k}(t)} -\ket{\varphi_{\psi_{k'}(t),{j^-}}}\right)\mathd t \nonumber\\
&- \sum_{k,k',j^-}\frac{N_{k'}(t)}{N} \left|\lambda_{\psi_{k'}(t),j^-}\right|\proj{\psi_k(t)} 
\delta\left(\ket{\psi_{k}(t)} -\ket{\varphi_{\psi_{k'}(t),{j^-}}}\right)\mathd t.
\end{align}
Now, second and third terms at the r.h.s. of the previous relation
provide the commutator and anti-commutator of the master equation we aim to,
while the eighth and ninth cancel each other, 
since the sum over $k$ in the latter term removes the $\delta$ and the remaining sum over $k'$ is of course equivalent to the sum over $k$ of the former term. On the other hand, 
the seventh and the last term sum up to
\begin{align}
&\sum_{k,j^+}\frac{N_k(t)}{N}
\lambda_{\psi_k(t),j^+}\proj{\varphi_{\psi_k(t),j^+}} \mathd t
- \sum_{k,k',j^-}\frac{N_{k'}(t)}{N} \left|\lambda_{\psi_{k'}(t),j^-}\right|\proj{\psi_k(t)} 
\delta\left(\ket{\psi_{k}(t)} -\ket{\varphi_{\psi_{k'}(t),{j^-}}}\right)\mathd t \nonumber\\
&= \sum_{k,j^+}\frac{N_k(t)}{N}
\lambda_{\psi_k(t),j^+}\proj{\varphi_{\psi_k(t),j^+}} \mathd t
- \sum_{k',j^-}\frac{N_{k'}(t)}{N} \left|\lambda_{\psi_{k'}(t),j^-}\right|\proj{\varphi_{\psi_{k'}(t),{j^-}}}
\mathd t \nonumber\\
&=\sum_k \frac{N_k(t)}{N} W_{\psi_k(t)}^J,
\end{align}
where in the last equality we used \reff{eq:psng2}. Then by virtue of the definition of $W_{\psi}^J$ we obtain also the term 
$\sum_{\alpha}c_\alpha(t)L_\alpha(t)(\sum_k\frac{N_k(t)}{N}\proj{\psi_k(t)})L_\alpha^\dag(t)=
\sum_{\alpha}c_\alpha(t)L_\alpha(t)\varrho(t) L_\alpha^\dag(t)$
of the master equation, while the terms depending on $\ell_{\psi_k(t),\alpha}$ in Eq.(\ref{eq:ell}) cancel out.


\begin{thebibliography}{39}
\bibitem{Breuer2002} H.-P. Breuer and F. Petruccione, \emph{The Theory of Open Quantum Systems}
(Oxford University Press, Oxford, 2002)
\bibitem{Rivas2012} {\'A}. Rivas and S. F. Huelga, \emph{Open Quantum Systems}
(Springer, New York, 2012)
\bibitem{Tanimura1989} Y. Tanimura and R. Kubo, J. Phys. Soc. Jpn. \bf{58}, 101 (1989); 
J.-J. Ding, R.-X. Xu, and Y. Yan, J. Chem. Phys. \bf{136}, 224103 (2012).
\bibitem{Macri1992} N. Makri, Chem. Phys. Lett. \bf{193}, 435 (1992).
\bibitem{Garraway1997} B. M. Garraway, Phys. Rev. A \bf{55}, 2290 (1997); 
L. Mazzola, S. Maniscalco, J. Piilo, K.-A. Suominen, and B. M. Garraway, Phys. Rev. A \bf{80}, 012104 (2009).
\bibitem{Breuer2008}
H.-P. Breuer and B. Vacchini, Phys. Rev. Lett. \bf{101}, 140402 (2008);
D. Chru{\'s}ci{\'n}ski and A. Kossakowski, Phys. Rev. A \bf{95}, 042131 (2017)
\bibitem{Prior2010} J. Prior, A. W. Chin, S. F. Huelga, and M. B. Plenio,
Phys. Rev. Lett. \bf{105}, 050404 (2010); 
D. Tamascelli, A. Smirne, J. Lim, S. F. Huelga, and M. B. Plenio, Phys. Rev. Lett. \bf{123}, 090402 (2019)
\bibitem{Zhang2012} W.-M. Zhang, P.-Y. Lo, H.-N. Xiong, M. Wei-Yuan Tu, and F. Nori,
Phys. Rev. Lett. \bf{109}, 170402 (2012).
\bibitem{Gualdi2013} G. Gualdi and C.P. Koch, Phys. Rev. A \bf{88}, 022122 (2013).
\bibitem{Ciccarello2013} F. Ciccarello, G. M. Palma, and V. Giovannetti, Phys. Rev. A \bf{87} 040103(R) (2013);
S. Lorenzo, F. Ciccarello, and G. M. Palma, Phys. Rev. A \bf{96} 032107 (2017).
\bibitem{Diosi2014} L. Di{\'o}si and L. Ferialdi, Phys.Rev.Lett. \bf{113} 200403 (2014);
L. Ferialdi, Phys.Rev.Lett. \bf{116} 120402 (2016).
\bibitem{Vacchini2016} B. Vacchini, Phys. Rev. Lett. \bf{117}, 230401 (2016); B. Vacchini, arXiv:1906.00693.
\bibitem{Tamascelli2018} D. Tamascelli, A. Smirne, S. F. Huelga, and M. B. Plenio,
Phys. Rev. Lett. \bf{120}, 030402 (2018);
A.D. Somoza, O. Marty, J. Lim, S. F. Huelga, M. B. Plenio, Phys. Rev. Lett. \bf{123}, 100502 (2019);
F. Mascherpa, A.Smirne, D. Tamascelli, P. Fernandez Acebal, S. Donadi, S. F. Huelga, M. B. Plenio, arXiv:1904.04822.
\bibitem{Strathearn2018} A. Strathearn, P. Kirton, D. Kilda, J. Keeling and B. W. Lovett,
Nat. Comm. {\bf 9}, 3322 (2018).
\bibitem{Lambert2019} N. Lambert, S. Ahmed, M. Cirio, F. Nori, Nat. Comm. \bf{10}, 3721 (2019).
\bibitem{Luchnikov2019} I. A. Luchnikov, S.V. Vintskevich, H. Ouerdane, and S. N. Filippov,
Phys. Rev. Lett. \bf{122}, 160401 (2019).
\bibitem{Carmichael1993} H.J. Carmichael, \emph{An Open System Approach to Quantum Optics,
Lectures Notes in Physics} (Springer, Berlin, 1993);
\bibitem{Gardiner2004}G.W. Gardiner and P. Zoller, \emph{Quantum Noise}  (Springer, Berlin, 2004).
\bibitem{Dalibard1992} J. Dalibard, Y. Castin, and K. M{\o}lmer, Phys. Rev. Lett. \bf{68}, 580 (1992).
\bibitem{Plenio1998} M. B. Plenio and P. L. Knight, Rev. Mod. Phys. \bf{70}, 101 (1998).
\bibitem{Basche1995}T. Basche, S. Kummer, and C. Brauchle, Nature \bf{373}, 132 (1995).
\bibitem{Peil1999}S. Peil and G. Gabrielse, Phys. Rev. Lett. \bf{83}, 1287 (1999).
\bibitem{Jelezko2002}F. Jelezko, I. Popa, A. Gruber, C. Tietz, J. Wrachtrup, A. Nizovtsev, and S. Kilin, Appl. Phys. Lett. \bf{81}, 2160 (2002); P. Neumann, J. Beck, M. Steiner, F. Rempp, H. Fedder, P. R. Hemmer, J. Wrachtrup, and F. Jelezko, Science \bf{329}, 542 (2010).
\bibitem{Gleyzes2007} S. Gleyzes, S. Kuhr, C. Guerlin, J. Bernu, S. Del{\'e}glise, U.B. Hoff, M. Brune, J.-M. Raimond, and S. Haroche, Natue \bf{446}, 297 (2007).
\bibitem{Vijay2011}R. Vijay, D. H. Slichter, and I. Siddiqi, Phys. Rev. Lett. \bf{106}, 110502 (2011)
\bibitem{Minev2019}Z. K. Minev, S. O. Mundhada, S. Shankar, P. Reinhold, R. Guti{\'e}rrez-J{\'a}uregui, R.J. Schoelkopf, M. Mirrahimi, H. J. Carmichael, and M. H. Devoret, Nature \bf{570}, 200 (2019).
\bibitem{Diosi1997} L. Di{\'o}si and W. T. Strunz, Phys. Lett. A \bf{235}, 569 (1997)
 L. Di{\'o}si, N. Gisin, and W. T. Strunz, Phys. Rev. A \bf{58}, 1699 (1998).
 \bibitem{Breuer2004} H.-P. Breuer, Phys. Rev. A \bf{70}, 012106 (2004).
\bibitem{Piilo2008} J. Piilo, S. Maniscalco, K. H{\"a}rk{\"o}nen, K.-A. Suominen, Phys. Rev. Lett. \bf{100} ,180402 (2008).
\bibitem{Barchielli2010} A. Barchielli, C. Pellegrini and F. Petruccione, Europh. Lett. \bf{91}, 24001 (2010);
A. Barchielli, P. Di Tella, C. Pellegrini, F. Petruccione, QP-PQ series \bf{27} 52 (2011).
\bibitem{Suess2014} D. Suess, A. Eisfeld, and W. T. Strunz, Phys. Rev. Lett. \bf{113}, 150403 (2014).
\bibitem{Gasbarri2018} G. Gasbarri and L. Ferialdi, Phys. Rev. A \bf{98}, 042111 (2018).
\bibitem{Megier2018} N. Megier, W.T. Strunz, C. Viviescas, and K. Luoma, Phys. Rev. Lett. \bf{120} 150402 (2018).
\bibitem{Breuer2009} H.-P. Breuer, E.-M. Laine, and J. Piilo, Phys. Rev. Lett. \bf{103}, 210401 (2009); H.-P. Breuer, E.-M. Laine, J. Piilo, and B. Vacchini, Rev. Mod. Phys. \bf{88}, 021002 (2016).
\bibitem{Rivas2010} {\'A}. Rivas, S. F. Huelga, and M. B. Plenio, Phys. Rev. Lett. \bf{105}, 050403 (2010); 
{\'A}. Rivas, S. F. Huelga, and M. B. Plenio, Rep. Prog. Phys. \bf{77}, 094001 (2014).
\bibitem{Li2018} L. Li, M.J. W. Hall, and H.M. Wiseman, Phys. Rep. \bf{759}, 1 (2018).
\bibitem{Piilo2009} J. Piilo, K. H{\"a}rk{\"o}nen, S. Maniscalco, and K.-A. Suominen, Phys. Rev. A \bf{79} 062112 (2009).
\bibitem{Gambetta2003} J. Gambetta and H.M. Wiseman, Phys. Rev. A \bf{68} 062104 (2003).
\bibitem{Diosi2008} L. Di{\'o}si, Phys.Rev.Lett. \bf{100} 080401 (2008); H.M. Wiseman
and J.M. Gambetta, Phys. Rev. Lett. \bf{101}, 140401 (2008).
\bibitem{Barchielli1991} A. Barchielli and V.P. Belavkin, J. Phys. A: Math. Gen. \bf{24} 1495 (1991).
\bibitem{Vacchini2011} B. Vacchini, A. Smirne, E.-M. Laine, J. Piilo, and H.-P. Breuer, New J. Phys. \bf{13}, 093004 (2011). 
\bibitem{Chruscinski2014} D. Chru{\'s}ci{\'n}ski and S. Maniscalco, Phys. Rev. Lett. \bf{112}, 120404 (2014).
\bibitem{Wissmann2015} S. Wi{\ss}mann, H.-P. Breuer, and B. Vacchini, Phys. Rev. A \bf{92}, 042108 (2015).
\bibitem{Cresser2010} J.D. Cresser, and C. Facer, Opt. Commun. \bf{283}, 773 (2010).
\bibitem{Hall2014} M.J.W. Hall, J.D. Cresser, L. Li, and E. Andersson, Phys. Rev. A \bf{89}, 042120 (2014).
\bibitem{Bernardes2015} N.K. Bernardes, A. Cuevas, A. Orieux, C.H. Monken, P. Mataloni, F. Sciarrino, and M.F. Santos, Sc. Rep. \bf{5} 17520 (2015).
\bibitem{Ferialdi2017} L. Ferialdi and A. Smirne, Phys. Rev. A \bf{96} 012109 (2017).
\bibitem{Megier2017} N. Megier, D. Chru{\'s}ci{\'n}ski, J. Piilo, and W.T. Strunz, Sci. Rep. \bf{7} 6379 (2017). 
\bibitem{Gorini1976} V. Gorini, A. Kossakowski, and E. C. G. Sudarshan, J. Math. Phys. \bf{17}, 821 (1976).
\bibitem{Laine2010} E.-M. Laine, J. Piilo, and H.-P. Breuer, Phys. Rev. A \bf{81} 062115 (2010).
\bibitem{Caiaffa2017} M. Caiaffa, A. Smirne, and A. Bassi, Phys. Rev. A \bf{95} 062101 (2017).
\bibitem{Diosi1986} L. Di{\'o}si, Phys. Lett. A \bf{114}, 451 (1986).
\bibitem{Diosi1988} L. Di{\'o}si, J. Phys. A \bf{21}, 2885 (1988); L. Di{\'o}si, J. Phys. A \bf{50}, 16LT01 (2017).
\bibitem{Gisin1990} N. Gisin, Helv. Phys. Acta \bf{63} 929 (1990).
\bibitem{Heinosaari2012} T. Heinosaari and M. Ziman, \emph{The Mathematical Language of Quantum Theory},
(Cambridge University Press, Cambridge, 2012)
\bibitem{Andersson2007} E. Andersson, J.D.  Cresser, and M.J.W. Hall, J. Mod. Opt. \bf{54}, 1695 (2007).
\bibitem{Chruscinski2015} D. Chru{\'s}ci{\'n}ski and F.A. Wudarski, Phys.Rev. A \bf{91}, 012104 (2015);
D. Chru{\'s}ci{\'n}ski and F.A. Wudarski, Phys.Lett. A \bf{377}, 1425 (2013).
\bibitem{Mitchison2018} M.T. Mitchison and M. B. Plenio, New J. Phys. \bf{20}, 033005 (2018).
\bibitem{Torre2018} G. Torre and F. Illuminati, Phys. Rev. A \bf{98} 012124 (2018). 
\bibitem{foot} In mathematical terms, this is due to the fact that the stochastic trajectories
are fixed by counting processes which are not Poisson processes, see Appendix \ref{app:cmi}. 

\end{thebibliography}
\end{document}